\documentclass[doublecol]{epl2}
\usepackage{graphicx}
\usepackage{bm}
\usepackage{amssymb}
\usepackage{amsmath}

\begin{document}

\title{Quantum Oscillations in Magnetic Field Induced Antiferromagnetic
Phase of Underdoped Cuprates : Application to Ortho-II YBa$_{\text{2}}$Cu$_{%
\text{3}}$O$_{\text{6.5}}$}
\shorttitle{Quantum Oscillations in Magnetic Field Induced Antiferromagnetic
Phase of Underdoped Cuprates }
\author{Wei-Qiang Chen\inst{1}, Kai-Yu Yang\inst{1}, T. M. Rice\inst{1,2}, \and F. C. Zhang\inst{1}}
\institute{
\inst{1} Department of Physics and Center of Theoretical and Computational Physics, University of Hong Kong, Hong Kong, China\\
\inst{2} Institut f\"{u}r Theoretische Physik, ETH-Z\"{u}rich, CH-8093 Z\"{u}rich, Switzerland }


\abstract{Magnetic field induced antiferromagnetic phase of the underdoped cuprates is
studied within the $t-t^{\prime }-J$ model. A magnetic field suppresses the
pairing amplitude, which in turn may induce antiferromagnetism. We apply our
theory to interpret the recently reported quantum oscillations in high
magnetic field in ortho-II YBa$_{\text{2}}$Cu$_{\text{3}}$O$_{\text{6.5}}$ \cite{dehaas} and propose that the total hole density abstracted from the
oscillation period is reduced by 50\% due to doubling of the unit cell.}

\date{today}
\maketitle

Recently, Doiron-Leyraud \textit{et al.} reported quantum oscillations
in the clean underdoped cuprate ortho-II YBa$_{\text{2}}$Cu$_{\text{3}}$O$_{%
\text{6.5}}$ in a high magnetic field at low temperatures \cite{dehaas} which imply
the presence of small Fermi surface (FS) pockets.
Their finding has shed new light on the nature of the anomalous spin gap
phase of underdoped cuprates, a key to understanding the mechanism of high $%
T_{c}$ superconductivity. The existence and topology of a FS \cite{H.Ding,ZXShen} is one of the  most important issues for underdoped
cuprates. Assuming the normal state is paramagnetic as in zero field,
angle resolved photoemission spectra on other underdoped
cuprates \cite{KMShen} imply a FS consisting of four hole pockets centered
on the Brillouin Zone (BZ) diagonals.

In this Letter we propose that the reported oscillations occur in an
antiferromagnetic (AF) ordered state, hence the oscillation period
corresponds to a Fermi pocket in the extended BZ. Our proposal is
substantiated by examining the interplay between AF and SC in a $t-t^{\prime
}-J$ model using a renormalized mean field theory (RMFT) \cite{ZhangGros88}.
A high magnetic field suppresses the pairing amplitude, which may in turn
induce AF ordering. The FS property is studied using a
phenomenological Green's function approach. The measured
oscillation period in the presence of AF ordering
implies a total hole concentration of 0.075 instead of
0.15 as interpreted by Doiron-Leyraud \textit{et al.} \cite{dehaas}. The
latter substantially exceeds the value of 0.10 obtained from an empirical
relation for $T_{c}=59K$ of this compound \cite{RXLiang-06} and a
theoretical estimate of 0.04 based on band structure calculations \cite%
{Bascones-PRB05}. The well ordered nature of this compound should enhance $%
T_{c}$, and the empirical formula should overestimate, not underestimate,
the hole concentration. We believe that our proposal gives a more reasonable hole density
for this compound.

The interplay of AF and SC order in underdoped cuprates has been a central
issue\cite{AFSC-exp,FengLee,Joynt,SO5,Ogata}. Recent
nuclear magnetic resonance data on multilayer cuprates have
established that uniform AF order can coexist with SC order in the same CuO$%
_{\text{2}}$ plane even in the absence of a magnetic field and have led
Mukuda \textit{et al.} to propose a revised phase diagram of the cuprates
with coexisting AF and SC order in the underdoped region \cite{Mukuda}.
Generally in cuprates AF and SC orderings are separated by a spin glass
region presumably due to their crystalline disorder. Coexistence of uniform
AF and SC order has been predicted from Variational Monte Carlo (VMC)
calculations on Gutzwiller projected wavefunctions \cite{Joynt, VMC-SC-AFM}.
We note that in the case of La$_{\text{2-x}}$Sr$_{\text{x}}$CuO$_{\text{4}}$
neutron scattering data in an applied magnetic field revealed subgaps in
spin excitations~\cite{Lake}, which were interpreted as evidence for a field
induced AF ordering in the SC state~\cite{Demler}. Our microscopic theory is
consistent with their phenomenological analyses~\cite{Demler}.

We consider the $t-t^{\prime }-J$ model on a square lattice \cite%
{Anderson87, ZhangRice},
\begin{equation}
H=-\sum_{i,j,\sigma }t_{ij}(c_{i\sigma }^{\dag }c_{j\sigma
}+h.c.)+J\sum_{\left\langle i,j\right\rangle }\mathbf{S}_{i}\cdot \mathbf{S}%
_{j}.  \label{Hamiltonian}
\end{equation}%
The hopping integrals are $t_{ij}=t$ for the nearest
neighbor (n.n.) pairs and $t_{ij}=t^{\prime }=-t/3$ for the next n.n. pairs.
$\mathbf{S}_{i}$ is a spin-1/2 operator, and the spin exchange
term ($J=t/3$) connects n.n. pairs. The constraint of no double
occupation at any lattice site is implied. We use a variational
wavefunction~\cite{FengLee,Joynt} to study the interplay of AF (or spin density wave (SDW))
and SC ordering,
\begin{align}
|\Psi \rangle  =P_{D}|\Psi _{0}\rangle, |\Psi _{0}\rangle =\prod_{\mathbf{k},s=\pm }\left( u_{\mathbf{k}s}+v_{%
\mathbf{k}s}d_{\mathbf{k}s\uparrow }^{\dag }d_{\mathbf{-k}s\downarrow
}^{\dag }\right) |0\rangle ,  \label{cowave}
\end{align}%
where the product of $\mathbf{k}$ runs over the reduced BZ, and $%
P_{D}=\prod_{i}(1-n_{i\uparrow }n_{i\downarrow })$ is the Gutzwiller
projection operator. $d_{\mathbf{k}s\sigma }$ is an annihilation operator of
the SDW quasiparticles, with $s=\pm $ for the upper or lower SDW bands,%
\begin{eqnarray}
d_{\mathbf{k}+,\sigma } &=&\cos {(\phi _{\mathbf{k}}/2)}c_{\mathbf{k}\sigma
}-\sigma \sin {(\phi _{\mathbf{k}}/2)}c_{\mathbf{k+Q}\sigma },  \label{SDWQP}
\\
d_{\mathbf{k}-,\sigma } &=&\sin {(\phi _{\mathbf{k}}/2)}c_{\mathbf{k}\sigma
}+\sigma \cos {(\phi _{\mathbf{k}}/2)}c_{\mathbf{k+Q}\sigma },  \notag
\end{eqnarray}%
where $\mathbf{Q}=\left( \pi ,\pi \right) $ is the AF wave vector.

To carry out the variation procedure we apply the Gutzwiller approximation
to replace the effect of the projection operator by a set of renormalization
factors, which are determined by statistical counting~\cite%
{ZhangGros88,Vanilla} and later improvement \cite{Ogata, Gan}. This
approximation or the renormalized mean field theory \cite{ZhangGros88}
incorporates the resonating valence bond physics proposed by Anderson \cite%
{Anderson87} and compares well with the VMC results \cite{Vanilla}.
Furthermore, as has been emphasized recently by Anderson \textit{et al.},
the RMFT describes many key features of the phase diagram \cite{Vanilla}.
Let $\langle \hat{O}\rangle $ be the expectation value of operator $\hat{O}$
in the projected state $|\Psi \rangle $, and $\langle \hat{O}\rangle _{0}$
be that in the corresponding unprojected state $|\Psi _{0}\rangle $, then
\begin{eqnarray}
\langle c_{i\sigma }^{\dag }c_{j\sigma }\rangle=g_{t,\sigma
}^{ij}\langle c_{i\sigma }^{\dag }c_{j\sigma }\rangle_{0},
\langle S_{i}^{\tau }S_{j}^{\tau }\rangle=g_{s}^{\tau }\langle
S_{i}^{\tau}S_{j}^{\tau}\rangle_{0}, \label{gfactor}
\end{eqnarray}%
where $\tau =x,y,z$, and $g$'s are the renormalization factors. In the AF
state, electron populations in the two sublattices $A$ (with net spin up)
and $B$ (with net spin down) are different. For the n.n. pair $(ij)$, $%
g_{t}=2\delta (1-\delta )/(1-\delta ^{2}+4m_{0}^{2})$. For the next n.n.
pairs on sublattice $A$ or $B$, $g_{t^{\prime }\sigma }^{A}=g_{t^{\prime
}-\sigma }^{B}=g_{t}(1+\delta +2\sigma m_{0})/(1+\delta -2\sigma m_{0})$. In
the above expressions, $\delta $ is the hole density, $m_{0}=\left\langle
S_{i}^{z}\left( -1\right) ^{i}\right\rangle _{0}$ is the staggered magnetic
moment in unit of $\mu _{B}$ in the state $|\Psi _{0}\rangle $. To address
the interplay between SC and AF ordering, we use an improved Gutzwiller
approximation of Ogata and Himeda for the spin exchange renormalization
factors, which takes the effect of intersite correlations into account \cite
{Ogata}. In their formalism, $g_{s}^{\tau }$ are also functions of $m_{0}$
and the pairing amplitude $\Delta $ in Eq. (\ref{kaidelta}), and are given
by $g_{s}^{x(y)}=g_{t}^{2}/\delta ^{2}(1+\eta _{1})^{-7}$, and $%
g_{s}^{z}=g_{s}^{x}(1+\eta _{2})$ with $\eta _{1},\,\eta _{2}$ both small
positive numbers, given in Ref.\cite{gs}. The difference between $g_{s}^{z}$
and $g_{s}^{x(y)}$ is due to the asymmetry in spin space.

Within the Gutzwiller approximation, varying a projected state
for $H$ in Eq. (\ref{Hamiltonian}) reduces to varying an
unprojected state $|\Psi _{0}\rangle $ for a renormalized  $H_{0}$,
\begin{equation}
H_{0}=-\sum_{i,j,\sigma }g_{t}^{ij}t_{ij}(c_{i\sigma }^{\dag }c_{j\sigma
}+h.c.)+J\sum_{\left\langle i,j\right\rangle ,\tau }g_{s}^{\tau }S_{i}^{\tau
}S_{j}^{\tau }.  \label{RMHamiltonian}
\end{equation}%
We introduce two mean fields, $\chi _{x}=\chi
_{y}=\chi $ for hopping and $\Delta _{x}=-\Delta _{y}=\Delta
$ for d-wave pairing,
\begin{align}
\chi =\sum_{\sigma }\left\langle c_{i+x \sigma }^{\dag }c_{i\sigma}
\right\rangle _{0},
\Delta _{x} =\left\langle c_{i+x\uparrow }c_{i\downarrow }-c_{i+x\downarrow
}c_{i\uparrow }\right\rangle _{0} \label{kaidelta}
\end{align}%
The singlet SC order parameter $\Delta _{SC}\approx \bar{g}_{t^{\prime
}}\Delta $, with $\bar{g}_{t^{\prime }}=(g_{t^{\prime }\uparrow
}^{A}+g_{t^{\prime }\uparrow }^{B})/2$. The pairing amplitude and the SDW
state described below define  $%
|\Psi _{0}\rangle $. We choose a standard SDW form,
\begin{equation}
\cos {(\phi _{\mathbf{k}})}=\epsilon _{\mathbf{k}}/\zeta _{\mathbf{k}%
},\,\,\zeta _{\mathbf{k}}=\sqrt{\epsilon _{\mathbf{k}}^{2}+\widetilde{\Delta
}_{m,\mathbf{k}}^{2}},  \label{SDWdispersion}
\end{equation}%
where $\epsilon _{\mathbf{k}}=-(2tg_{t}+3J\bar{g}_{s}\chi /4)\gamma _{+}(%
\mathbf{k})$ is the kinetic energy contribution of n.n. hopping including a
self-energy term of $\chi $, with $\bar{g}_{s}=(g_{s}^{z}+2g_{s}^{xy})/3$,
and $\widetilde{\Delta }_{m,\mathbf{k}}=m_{v}+2t^{\prime }(g_{t^{\prime
}\uparrow }^{A}-g_{t^{\prime }\uparrow }^{B})\theta _{\mathbf{k}}$. $m_{v}$
is a variational parameter which determines $m_{0}$, $\gamma _{\pm }(\mathbf{%
k})=\cos k_{x}\pm \cos k_{y}$, and $\theta _{\mathbf{k}}=\cos k_{x}\cos
k_{y} $. The d-wave pairing variational functions are chosen as
\begin{align}
v_{\mathbf{k}\pm }^{2}& =(1-\xi _{\mathbf{k}\pm }/E_{\mathbf{k}\pm })/2,
\label{uvfactor} \\
u_{\mathbf{k}\pm }v_{\mathbf{k}\pm }& =\Delta _{v}(\mathbf{k})/2E_{\mathbf{k}%
\pm },  \notag \\
\xi _{\mathbf{k}\pm }& =\pm \zeta _{\mathbf{k}}-4t^{\prime }\bar{g}%
_{t^{\prime }}\theta _{\mathbf{k}}-\mu ,  \notag \\
\Delta _{v}\left( \mathbf{k}\right) & =(3/4)J\bar{g}_{s}\Delta _{v0}\gamma
_{-}(\mathbf{k}),  \notag
\end{align}%
with $E_{\mathbf{k}\pm }=\sqrt{\xi _{\mathbf{k}\pm }^{2}+\Delta _{v}(\mathbf{%
k})^{2}}$, and $\Delta _{v0}$ is a variational parameter. Note that because
of the mean field dependence of the $g$-factors here, $\Delta _{v0}\neq
\Delta $ in general.

\begin{figure}[tbph]
\centerline{\resizebox{80mm}{!}{\includegraphics[]{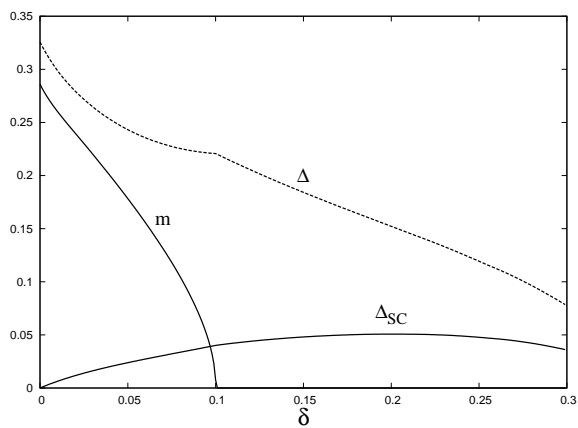}}}
\caption{Pairing mean field $\Delta $ (dashed line), superconducting order
parameter $\Delta _{SC}$ and antiferromagnetic staggered magnetization $m$
(solid lines) as functions of hole concentration $\protect\delta $ for $%
t-t^{\prime }-J$ model, with $J=t/3$, $t^{\prime }/t=-1/3$.}
\label{fig:pd}
\end{figure}

The variational energy per site $E=\left\langle H\right\rangle /N_{s}$ can
then be expressed in terms of the mean field parameters,
\begin{eqnarray}
E &=&-4g_{t}t\chi -8t^{\prime }\bar{g}_{t^{\prime }}\chi _{1}+4t^{\prime
}(g_{t^{\prime }\uparrow }^{A}-g_{t^{\prime }\uparrow }^{B})\chi _{2}
\label{energy} \\
&&-(3\bar{g}_{s}/4)J(\Delta ^{2}+\chi ^{2})-2g_{z}Jm_{0}^{2},  \notag
\end{eqnarray}
where $\chi _{1}=-(1/2N_s)\sum_{\mathbf{k},s}\xi _{\mathbf{k}%
s}\theta _{\mathbf{k}}/E_{\mathbf{k}s}$, $\chi _{2}=-(1/2N_s)\sum_{%
\mathbf{k}s}s\widetilde{\Delta }_{m,\mathbf{k}}\xi _{\mathbf{k}s}\theta _{%
\mathbf{k}}/\zeta _{\mathbf{k}}E_{\mathbf{k}s}$. The variational ground
state is obtained by minimizing $E$ with respect to $m_{v}$, $\Delta _{v0}$,
$\chi $, $\mu $, under constraint of the constant total hole concentration $%
\delta $. In Fig. \ref{fig:pd}, we plot the pairing mean field $\Delta $,
the SC order parameter $\Delta _{SC}$, and the staggered magnetization $m=%
\sqrt{g_{s}^{z}}m_{0}$ in the projected state $|\Psi \rangle $, as functions
of doping $\delta $. At $\delta <\delta _{c}=0.1$, AF order co-exists with
SC order; at $\delta >\delta _{c}$, the ground state has only SC order with $%
m=0$. This result is similar to the variational Monte Carlo calculation
which gives $\delta _{c}\simeq 0.10$ \ for the $t-J$ model \cite{VMC-SC-AFM}.

We now consider the effect of an external magnetic field on the phase
diagram. Cuprates are type II superconductors, and a magnetic field
penetrates in the form of vortices. The SC order outside
the vortex cores is substantially suppressed if the applied field $%
H_{app}$ is comparable with the critical field $H_{c2}$ as in the experiments in Ref. [1],
but quantum oscillations may still occur \cite{maniv}

\begin{figure}[tbph]
\centerline{\resizebox{80mm}{!}{\includegraphics[]{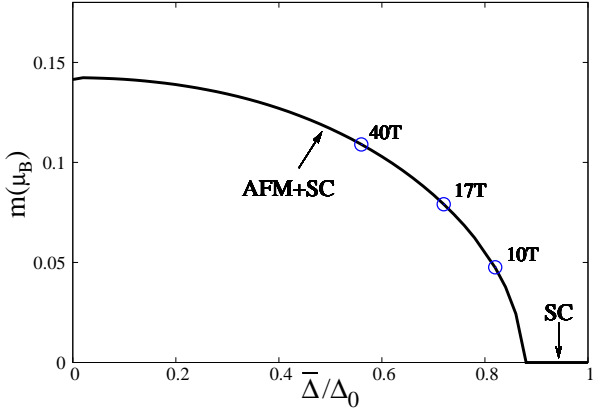}}}
\caption{Staggered magnetization $m$ obtained from Eq. (\protect\ref{energy}%
) of RMFT as a function of the ratio of the suppressed pairing amplitude and
the zero-field pairing amplitude $\protect\alpha =\overline{\Delta }/\Delta
_{0}$ at doping $\protect\delta =1.2\protect\delta _{c}$ ($\protect\delta %
_{c}=0.10$) ($t=0.3$ eV, $J=0.1$ eV, $t^{\prime }=-0.1$ eV), where $%
\overline{\Delta }$ is the spatially averaged pairing mean field $\Delta
\left( \mathbf{r}\right) $ as discussed in context, $\Delta _{0}$ is the
pairing amplitude at zero field. Circles indicate the estimated
corresponding applied fields (see the text).}
\label{fig:mag}
\end{figure}

We use the RMFT to show that the suppression of the SC pairing
amplitude due to the magnetic field may induce AF ordering. Consider the
case $\delta \ge \delta _{c}$, so that the ground state at $H_{app}=0$ is a
pure SC state with the values in Eq. (\ref{kaidelta})  $
\Delta _{0}$ and $m_{0}=0$. We approximate the effect of $H_{app}$ on the
background SC region by a suppression of the pairing amplitude from $\Delta
_{0}$ to an average value $\overline{\Delta }=\alpha \Delta _{0}$, with $%
\overline{\Delta }$ estimated below. Then Eq. (\ref{energy}) allows us to
carry out the variation for a given $\Delta $ or equivalently for a fixed
value of the variational parameter $\Delta _{v0}$. We have found that at
 $\delta  > \delta _{c}$, a suppression of $\Delta _{SC}$ and
hence $\Delta $ will induce AF ordering $m\neq 0$. In Fig. \ref{fig:mag} we
plot $m$ as a function of $\alpha $ for $\delta =1.2\delta _{c}=0.12$. $%
\overline{\Delta }$ may be estimated as follows. Let $\xi _{coh}$ be the SC
coherence length, and $d$ the average distance of two neighboring vortices.
We have $d/\xi _{coh}\approx \sqrt{H_{app}/H_{c2}}$, and $\overline{\Delta }%
=\pi ^{-1}(d/2)^{-2}\int_{r\leq d/2}\Delta \left( \mathbf{r}\right) d\mathbf{%
r}$, where $\Delta \left( \mathbf{r}\right) $ is the shape of the pairing
mean field around a vortex. For a rough estimate, we use for
s-wave the form, $\Delta \left( \mathbf{r}\right) =\Delta _{0}\tanh \left( r/\xi
\right) $ \cite{Gygi}. The value of $H_{c2}$ of underdoped high Tc
superconductors is about $160T$ \cite{Ong}, so the suppression ratio $\alpha
$ at $H_{app}=10T$ $(\sim H_{c2}/16)$, $18T$ $(\sim H_{c2}/9)$, and $40T$ $
(\sim H_{c2}/4)$ are $0.82$, $0.72$, and $0.56$, respectively. The points
corresponding to these values are denoted in Fig.~\ref{fig:mag} with open
circles.

We turn now to the implications of our calculations for the interpretation
of the recent experiments by Doiron-Leyraud \textit{et al.}~\cite{dehaas}.
In their experiments, the Hall resistance of underdoped YBa$_{\text{2}}$Cu$_{%
\text{3}}$O$_{\text{6.5}}$, as a function of magnetic field, shows clear
oscillation in the vortex liquid phase. The FS area thus obtained
is found by using Onsager relation and corresponds to small FS
pockets with a charge carrier density of 0.0375. The compound is SC without
AF ordering at low temperature in the absence of magnetic field. Estimates
of the hole doping from the SC transition temperature give a value $\delta=0.1$.
Since Fermi arcs have been observed in angle resolved photoemission spectra
around the nodal points in Na-CCOC at similar hole doping~\cite{KMShen},
Doiron-Leyraud \textit{et al.} interpreted the full FS to be
composed of 4 such Fermi pockets. Here we point out that while the low
temperature phase at $H_{app}=0$ is SC, the high magnetic field may induce
AF ordering as discussed above. If this is the case, the area enclosed by
the full FS will be double the pocket area. Thus the quantum
oscillation may actually indicate a hole doping of only 0.075 instead of
0.15.

\begin{figure}[tbp]
\includegraphics [width=8.0cm,height=8.0cm] {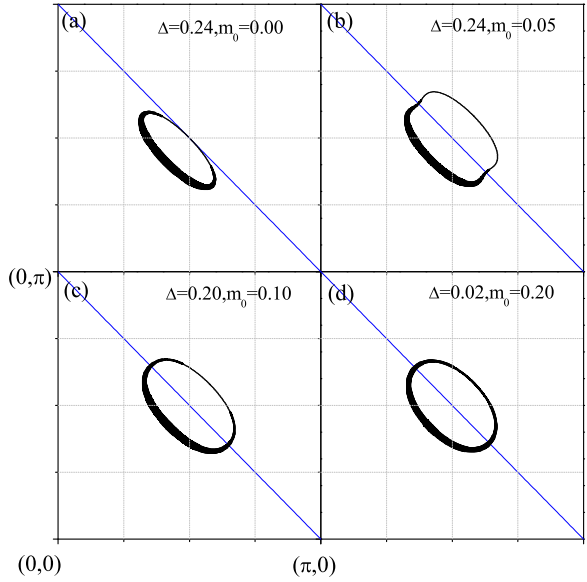}
\caption{Evolution of Fermi surface in one quadrant of BZ calculated from
Eq. (\protect\ref{Green}-\protect\ref{LSR}) at hole doping $\protect\delta %
=0.075$. The period of quantum oscillations is given by the area of the
closed Fermi pocket, which is $\protect\delta /4$ for (a) in the absence of
AF order and is $\protect\delta /2$ for (b)-(d) in the AF ordered state. The
thickness of the curves represents the spectral weight.}
\label{fig:Fermi_surface}
\end{figure}

Since quantum oscillations in the vortex liquid phase are known to reproduce
those of the underlying normal state \cite{maniv}, we examine the
FS pockets in the non-SC state. To this end we extend the
phenomenological theory we developed earlier for the pseudogap state \cite%
{KYYang-prb-06} to include AF ordering. We make an ansatz for the coherent
part of the Green function (GF) which represents a quasiparticle,
\begin{equation}
G_{AF}^{s}(\mathbf{k},\omega )=\frac{z}{\omega -\xi _{\mathbf{k}s}-\Sigma
_{s}(\mathbf{k},\omega )},  \label{Green}
\end{equation}%
where $z$ is a numerical factor for the weight of the coherent part. The
self energy takes the form
\begin{equation}
\Sigma _{s}(\mathbf{k},\omega )=\Delta ^{2}\gamma _{-}^{2}(\mathbf{k}%
)/(\omega +s\zeta _{\mathbf{k}}),  \label{selfenergy}
\end{equation}%
where $\zeta _{\mathbf{k}s}$ and $\xi _{\mathbf{k}s}$ are given in Eqs. (\ref%
{SDWdispersion}) and (\ref{uvfactor}) with the replacement of $m_{v}$ by $%
m_{0}$. The Luttinger sum rule \cite{KRT,AGD} for the number of electrons
can be formulated as
\begin{equation}
N_{e}=\sum_{\mathbf{k}s}\Theta \left[ ReG_{AF}^{s}(\mathbf{k},\omega =0)%
\right] ,  \label{LSR}
\end{equation}%
where $\Theta (x)$ is the Heaviside step function. Note that the boundary of
the reduced BZ in the SDW state coincides with the Luttinger surface of
zeroes of the corresponding GF in the absence of AF ordering.
The FS of $G_{AF}^{s}$ depends on the staggered magnetization $%
m_{0}$. In Fig. \ref{fig:Fermi_surface}, we show the FS for
several values of $m_{0}$ at a hole doping $\delta =0.075$. In that case,
the upper SDW band is completely empty and the lower SDW band is partially
filled. $\Delta $ is chosen to be the value defined as in Eq.(\ref{kaidelta}%
) that gives the lowest energy in the RMFT
for a given $m_{0}$. In such case the orbits of a quasiparticle in a
magnetic field are confined to the reduced BZ and consist of two ellipses
each composed of two half ellipses centered on opposite diagonals BZ. The
period of quantum oscillations corresponds to one half of the total hole
density. In the absence of AF order, \textit{i.e.} $m_{0}=0$, the
phenomenological GF has now four separate pockets near the
nodal points and associating the period of the quantum oscillations with the
area of a pocket leads to a total hole density twice as large.

The electron GF can be obtained from the SDW GF. The resulting
electron spectra weights relevant for angle resolved photoemission
spectra measurements are shown in Fig. \ref
{fig:Fermi_surface} by the width on the Fermi pockets. The spectral weight
distribution inside the reduced zone at small $m_{0}$ is similar to
our phenomenological theory for the pseudogap state \cite{KYYang-prb-06},
\textit{i.e.} a substantial weight only on the inner edge. The spectral
weight outside the reduced BZ is very small but grows as $m_0$ increases.
With increasing $m_{0}$, the
pockets become more ellipsoidal.

In summary, we have used the renormalized mean field theory to show that a
high magnetic field may induce antiferromagnetism in underdoped
superconducting state. Applying our theory to the recently observed quantum
oscillations in a magnetic field in underdoped cuprates ~\cite{dehaas}, we
propose that the period measured in oscillations implies a hole density of
0.075, instead of 0.15 as originally proposed.

Since the original manuscript was prepared, quantum oscillations have been observed in
$YBa_2Cu_4O_8$ \cite{yelland}. Again assuming 2 rather than 4 pockets gives a more reasonable
value of the hole density $\delta=0.1$. Lee has drawn our attention to his proposal of a doubled unit cell
due to a staggered flux phase \cite{lee2007} as an alternative to the AF phase considered here.
We have carried out a similar analysis for this case and found
that it has a slightly lower energy than the AF phase. It may well be possible to observe the broken
symmetry in moderate fields (eg see Fig. 2) and so check for both possibilities experimentally.
Renormalized mean field theory show that a
high magnetic field may induce antiferromagnetism in underdoped
superconducting state. Applying our theory to the recently observed quantum
oscillations in a magnetic field in underdoped cuprates ~\cite{dehaas}, we
propose that the period measured in oscillations implies a hole density of
0.075, instead of 0.15 as originally proposed.

 After we submitted the paper, LeBoeuf \textit{et al.}
\cite{Leboeuf} published their new data which strongly suggest
that at high magnetic fields in both YBCO compounds an unpaired
normal state is stabilized with a longer period AF or other
superlattice rather than the simple AF or flux ordering that we
consider in this paper. This implies that the overall negative
sign of the Hall signal cannot be attributed to a contribution
from vortex motion and is due rather to electron than to hole pockets.
Millis and Norman have found that such an electron pocket may
appear and dominate magnetotransport in the presence of an AF
superlattice with an 8-fold period perpendicular to the chains at
1/8 doping.\cite{Millis} Further calculations extending our
results are underway and will be reported elsewhere. 

\acknowledgments We thank L. Taillefer for providing us with a
copy of their paper prior to publication and we thank him and P.
A. Lee for discussions. This work was in part supported by RGC
grant of HKSAR. FCZ wishes to thank Joint Theory Institute of
University of Chicago and Argonne National Laboratory for its
hospitality, where part of the work was carried out. Support by
the MANEP program of the Swiss National Foundation and a Visiting
Professorship at the University of Hong Kong is also gratefully
acknowledged.

\begin{thebibliography}{99}
\bibitem{gs} The small parameters $\eta_1$ and $\eta_2$ in $g_s^{x(y)} $ and $g_s^z$ are given by $\eta_1= 2 X/(1 - \delta^2 + 4m_0^2)^2$, $\eta_2=8 m_0^2 (2X_1+X_1^2)/(8m_0^2+X_2)$, $X = \delta^2 \left(\Delta^2 - \chi^2 \right) + 4 m_0^2 X_2 + X_2^2 / 4$, $X_2 = \chi^2 + \Delta^2 $, and $ X_1=3X_2(1-\delta)^2/[(1-\delta^2 +4m_0^2)(1+\eta_1)^3]$.
\end{thebibliography}
\end{document}